\documentclass[journal=nanolett, manuscript=article]{achemso}

\usepackage{chemformula} 
\usepackage[T1]{fontenc} 
\usepackage{siunitx}
\usepackage{mathtools}
\usepackage{xcolor}

\usepackage{verbatim}

\setkeys{acs}{maxauthors = 0}

\author{Allison M. Green}
\affiliation[UTChemE]{McKetta Department of Chemical Engineering, University of Texas at Austin, Austin, Texas 78712, United States}
\author{Charles K. Ofosu}
\affiliation[UTChem]{Department of Chemistry, University of Texas at Austin, Austin, Texas 78712, United States}
\author{Jiho Kang}
\affiliation[UTChemE]{McKetta Department of Chemical Engineering, University of Texas at Austin, Austin, Texas 78712, United States}
\author{Eric V. Anslyn}
\affiliation[UTChem]{Department of Chemistry, University of Texas at Austin, Austin, Texas 78712, United States}
\email{anslyn@austin.utexas.edu}
\author{Thomas M. Truskett}
\affiliation[UTChemE]{McKetta Department of Chemical Engineering, University of Texas at Austin, Austin, Texas 78712, United States}
\alsoaffiliation[UTPhys]{Department of Physics, University of Texas at Austin, Austin, Texas 78712, United States}
\email{truskett@che.utexas.edu}
\author{Delia J. Milliron}
\affiliation[UTChemE]{McKetta Department of Chemical Engineering, University of Texas at Austin, Austin, Texas 78712, United States}
\alsoaffiliation[UTChem]{Department of Chemistry, University of Texas at Austin, Austin, Texas 78712, United States}
\email{milliron@che.utexas.edu}

\title{Assembling Inorganic Nanocrystal Gels}

\begin{document}

\begin{tocentry}
\includegraphics[width=\textwidth]{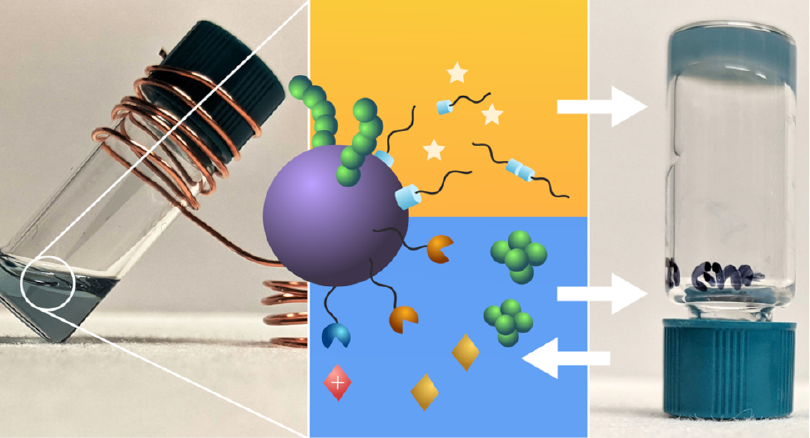}
\end{tocentry}

\begin{abstract}
Inorganic nanocrystal gels retain distinct properties of individual nanocrystals while offering tunable, network structure-dependent characteristics. We review different mechanisms for assembling  gels from colloidal nanocrystals including (1) controlled destabilization, (2) direct bridging, (3) depletion, as well as linking mediated by (4) coordination bonding or (5) dynamic covalent bonding, and we highlight how each impacts gel properties. These approaches use nanocrystal surface chemistry or the addition of small molecules to mediate inter-nanocrystal attractions. Each method offers advantages in terms of gel stability, reversibility, or tunability and presents new opportunities for design of reconfigurable materials and fueled assemblies.

\end{abstract}


\section*{Introduction}
Colloidal inorganic nanocrystals (NCs) have distinctive, synthetically tunable properties including plasmonic response, electronic conductivity, and catalytic activity.\cite{Talapin2010, Agrawal2018} These properties can be incorporated into extended NC structures such as superlattices\cite{Boles2016,Murray1995} and gels \cite{Matter2020, Niederberger2017} and augmented by structure-dependent interactions. Gels in particular are characterized by porous networks of interconnected NCs and offer reversible and continuous control over the number of NC neighbors, influencing coupling and collective properties important for applications in catalysis, optics, electrochromics, and energy storage.\cite{Liu2015B, Hewa-Rahinduwage2021} Although strategies for forming gels of microscale colloids are well developed,\cite{Zaccarelli2007} the applicability and effectiveness of those approaches for assembling nanoscale colloidal gels are still being established.\cite{Sherman2021} An understanding of how to tune the effective interactions between NCs and their equilibrium phase behavior is essential for designing NC gel assemblies with desired structure and characteristics.

Colloidal stability of dispersed, as-synthesized NCs is typically provided by inter-NC repulsive forces due to ligands bound to their surfaces. Gelation strategies can involve either weakening these stabilizing repulsions, introducing inter-NC attractions, or both.\cite{Matter2020} The resulting NC aggregation kinetics is influenced by the strength and range of the resulting interactions. In the limit where the repulsive forces are negligible, NCs stick to one another upon collision, and the rate of cluster formation is determined by the NC diffusivity, i.e., diffusion limited cluster aggregation (DLCA) occurs.\cite{Jungblut2020}
When sufficient interparticle repulsive interactions are present, diffusion occurs much faster than the binding of particles or clusters, leading to reaction limited cluster aggregation (RLCA). Early examples of nanoparticle (NP) gelation involved sol-gel chemistry, where gels were formed by dispersing NPs in a silica sol to effectively ``glue'' the network together.\cite{Morris1999} This laid the groundwork for and motivated further NC gelation studies due to the promising catalytic properties these gels exhibited. Subsequently, noble metal NP gels were formed by spontaneous aggregation of unstable particles in a single step reduction of metal precursors without first stabilizing the particles.\cite{Liu2012,Liu2015B} However, forming stable NC dispersions is a prerequisite to forming gels with properties that can be understood and designed based upon the characteristics of their NC building blocks. Herein, we examine routes to assemble NC gels and discuss how they lead to assemblies with distinctive properties.

Broadly speaking, NC gelation strategies are based on (1) NC surface chemistry or (2) introduction of small molecules to mediate inter-NC attractions (Fig. 1). The former depends more sensitively on the NC building block and relies on chemical modification of the NC surface to introduce destabilizing attractions. Examples include progressive ligand removal,\cite{Mohanan2004} or bridging particles using linkers that directly bond with NC surfaces \cite{Song2020}. The \emph{simplicity} of these chemistries makes such routes straightforward to implement while still offering a rich variety of assemblies. On the other hand, the latter requires the design of secondary molecules to act as linkers that bridge the surface ligands on different NCs\cite{Singh2015,Dominguez2020} or as depletants to promote entropic attractions.\cite{Cabezas2020} The use of small-molecules effectively decouples NC synthesis from gelation; such \emph{modularity} enables universal gelation routes that are independent of NC composition.

This review is organized into sections by gelation mechanism. Our goal is to highlight key examples of each mechanism and describe how the modularity or simplicity of their design impacts the resulting characteristics of NC gels (Fig. 1). We aim to establish a clear connection between each mechanism, and the resulting gels to guide judicious selection of strategies and NC building blocks. With a better understanding of the advantages of each strategy follows new opportunities to engineer functional materials.

\begin{figure}
    \centering
    \includegraphics[width=0.8\textwidth]{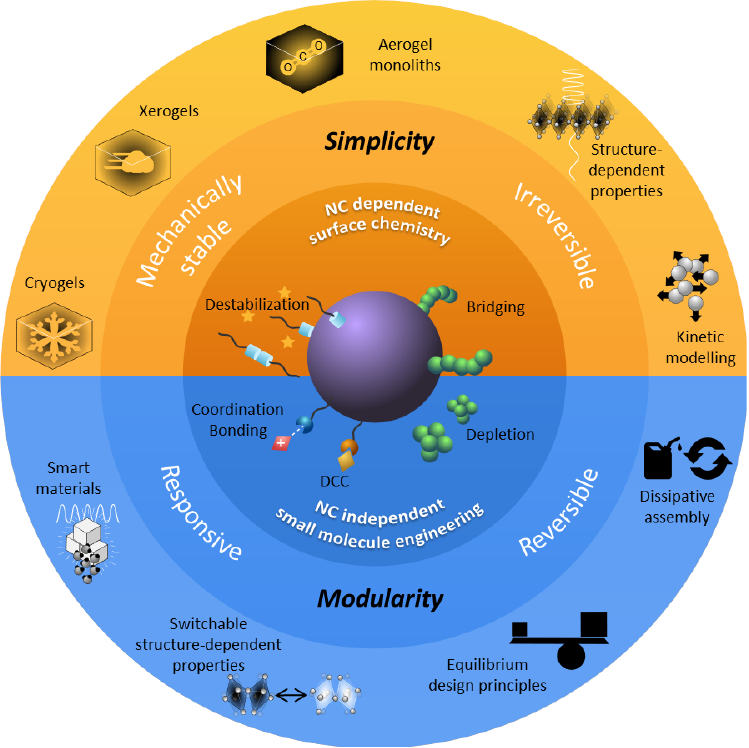}
    \caption{  Schematic overview of (inner circle) NC gelation mechanisms highlighted in this review, (middle circle) characteristics of resulting gels, and (outer circle) applications or properties accessible based on the chosen gelation route.\label{fig:}}
\end{figure}

\section*{Controlled aggregation by progressive destabilization}

Progressive destabilization via ligand removal is a straightforward route to creating robust and stable gel networks that can be further transformed into aerogels or xerogels.\cite{Arachchige2007B} Removing ligands weakens the short-range repulsive interactions that hold the NCs apart, enabling van der Waals (vdW) attractions to dominate and induce aggregation, often leading to the formation of attractions that strongly bind one NC directly to the next (Fig. 2a). However, if not well controlled, such destabilization can lead to irreversible formation of dense aggregates within the gels, paired with a total loss of the properties exhibited by the original NC building blocks.

Chemical and photo-oxidation have been commonly employed to drive desorption of surface ligands by converting their surface-binding domain, ultimately inducing aggregation of NCs. For example, CdS NCs functionalized by 4-fluorophenylthiol ligands were transformed into gels via chemical oxidation.\cite{Gacoin2001} Upon the addition of oxidant (\ch{H2O2}) to a NC dispersion, the bound ligands sterically stabilizing NCs were gradually desorbed as dithiol and fluorothiol species. The strength of inter-NC repulsions was tuned by varying the molar ratio of \ch{H2O2} to ligands. Below a minimum oxidant-ligand ratio, the NCs remained dispersed due to stabilization against aggregation from the bound ligands. Above a threshold, however, enough ligands were removed to enable vdW attractions to prevail, inducing the irreversible formation of gels. Additional oxidant beyond the gelation threshold first led to the shrinkage of gels and eventually to precipitation. Similarly, photooxidation has been used to induce aggregation of thiol-coated NCs.\cite{Aldana2001} In the presence of light and oxygen, bound thiol ligands are photocatalytically oxidized into disulfides and desorbed from NC surfaces with the NCs acting as the photocatalyst. Overall, these oxidation methods require an appropriate pairing of the ligand chemistry and the destabilizer to effectively form gels.

Another robust platform to progressively destabilize NCs into gels is via electrooxidation or ``electrogelation.'' This approach was first demonstrated with CdS quantum dots as a means to interface chalcogenide gels to electrodes.\cite{Hewa-Rahinduwage2020} Studies revealed a three-step mechanism: (1) electrochemical removal of thiolate ligands via cleavage of the Cd-S bond to form dithioglycolates, (2) solvation and subsequent detachment of the exposed Cd ions on the surface of the quantum dot, producing a S-rich surface and (3) electrochemical crosslinkage of the quantum dots by dichalcogenide bond formation. As electrogelation relies on electrochemical reactions at the NP surface, ligand exchange of the long-chain surfactants stabilizing the NPs with shorter chain ligands facilitated gelation. Subsequent reversal of the electrode potential led to dissassembly.\cite{Hewa-Rahinduwage2020}  Similarly, \ch{Pb_xCd_{1-x}Se} gels were formed via electrogelation of CdSe quantum dots followed by cation exchange and exhibited high \ch{NO_2} gas-sensing performance at an optimal ratio of Pb:Cd.\cite{Geng2021}

The high structural stability of solvogels formed by progressive destabilization routes offers the opportunity to further process the gels into xerogels and aerogels (Fig. 2b).\cite{Mohanan2005,Arachchige2007A} Aerogels are formed by removing the solvent in a solvogel via supercritical drying. \cite{Husing1998} Due to the absence of surface tension in supercritical solvent, aerogels can maintain the structural integrity of the original framework, avoiding pore collapse during drying. Chalcogenide NC aerogels are typically prepared via a three step process: synthesis of thiolate-capped NCs, formation of solvogels via chemical- or photo-oxidation, and, lastly, removing solvent from the wet gel by supercritical carbon dioxide drying. Although photooxidation generally shows slower gelation kinetics compared to chemical oxidation, aerogels formed by either oxidation method display similar material properties.\cite{Arachchige2006} Xerogels, on the other hand, are formed by drying solvogels under ambient conditions. These gels exhibit denser structures than the original solvogel because capillary forces cause their pores to partially collapse. Both drying processes have been used to create porous monoliths with tunable optical properties; the resulting gels exhibit optical responses distinct from those either NC dispersions or densely packed NC solids due to their structural differences (Fig. 2b).\cite{Arachchige2007A,Arachchige2007B}

Much like chalcogenides, noble metal NPs (i.e. Pt, Au, Ag) can also be transformed into aerogels; these are of significant interest due to their promising catalytic activity, specifically for ethanol oxidation and the oxygen reduction reaction \cite{Liu2012}. Metallic NP gels are typically obtained by adding destabilizers to a concentrated NP dispersion. \cite{Liu2015B,Bigall2009} For example, dopamine was employed to destabilize Au NPs into hydrogels by displacing their stabilizing ligands;\cite{Wen2016} dopamine can form complexes with ligands, such as beta-cyclodexrin, as well as adsorb onto the NP surface itself. Higher dopamine concentrations led to faster gelation kinetics. Subsequently, aerogels with a broad pore size distribution were formed via supercritical drying. Cyclic voltammetry and amperometry were used to evaluate the catalytic performance of Au aerogels with different functional groups: beta-cyclodexrin ($Au_{\beta-CD}$), citrate ($Au_{Cit}$), and non-stabilizers ($Au_{NS}$) (Fig. 2c). The aerogels exhibited small cyclic voltammetry peak potential separations, indicating high electron transfer efficiency, and high amperometric sensitivity to glucose concentration. The high catalytic activity of the aerogels can be attributed to their porous structure with large electrochemically active surface areas.

Controlled destabilization of NCs into gel networks relies upon the NC surface chemistry and selection of effective destabilizers. Controlling the structure of these solvogels can be challenging as they are non-equilibrium structures that can continuously and irreversibly change or densify with time.\cite{Lu2008} Furthermore, fusion of metal nanoparticles can lead to a loss or weakening of plasmonic optical absorption.\cite{Wen2016} Nonetheless, gels formed via this mechanism have commonly been transformed into highly stable aerogel monoliths \cite{Mohanan2004,Arachchige2007A}. These are relevant for (electro)catalysis due to their large surface areas, as well as for sensing and photoelectrochemical cells due to increased diffusion rates of small molecules (i.e. ethanol, glucose) through their pores.\cite{Arachchige2006,Liu2015B} Thus, while there are some limitations, the robust aerogels formed via this route offer a promising approach for catalysis, sensing, and for example, in fuel cell applications.

\begin{figure}
    \centering
    \includegraphics[width=1\textwidth]{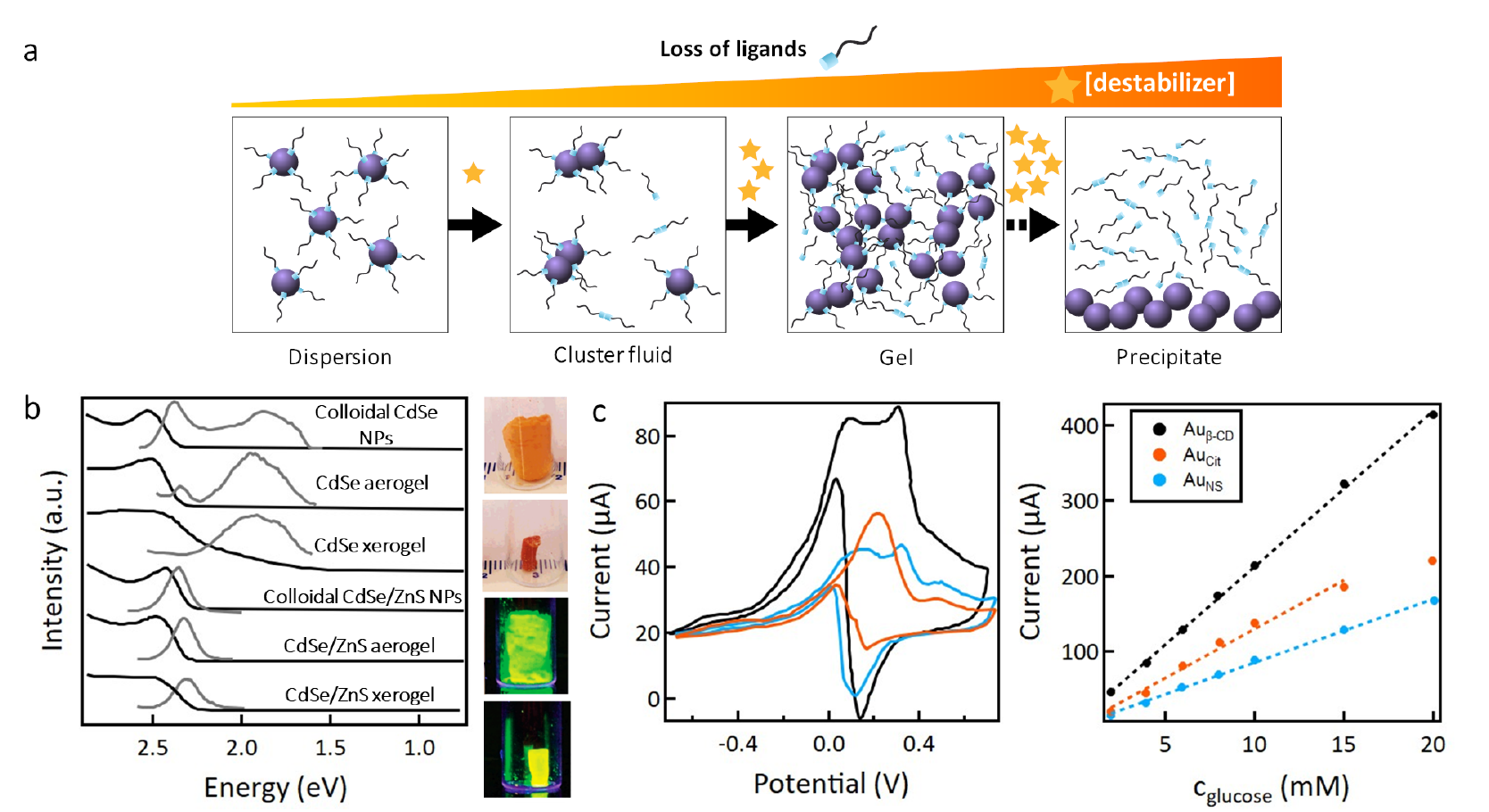}
    \caption{  (a) Schematic of NC controlled aggregation by progressive destabilization into gels, with increasing concentration of destabilizer. (b) Left: Optical absorption (black) and photoluminescence (gray) spectra for CdSe and CdSe/ZnS dispersions, aerogels, and xerogels. Right: Digital photographs of a NC aerogel and xerogel, under ambient light (top), and UV light (bottom). (c) Electrochemical testing of Au NP aerogel-modified electrodes. Left: Cyclic voltammetry for glucose oxidation. Right: Amperometry calibration curves of electrocatalytic current vs. glucose concentration to evaluate aerogel glucose oxidation activity (electrocatalytic current) and sensitivity (change in current per change in glucose concentration for a given electrode surface area). Adapted with permission from:(a) ref.~\citenum{Gacoin2001}, Copyright 2001 American Chemical Society. (b) ref.~\citenum{Arachchige2007A}, Copyright 2007 American Chemical Society. (c) ref.~\citenum{Wen2016}, Copyright 2016 American Chemical Society. \label{fig:}}
\end{figure}

\section*{Direct bridging}
Another strategy to assemble NCs into gels is to directly connect NCs via a secondary molecule, relying on their intrinsic surface chemistry. The NCs are connected surface-to-surface, as in the previous section; however, this is achieved by adding a macromolecule to bridge the particles. This effectively limits particle valence (the number of neighbors bonded to each NC) by the number of bridges introduced, offering additional control over the assembly. Ligand bridges have been used to assemble NCs into supercrystals by exploiting ligand crosslinking reactions or modulating ligand-solvent interactions \cite{Dreyer2016, Domenech2019}. Because the bridging molecule must either adsorb or bind to the surface of the NC itself, it must be selected in consideration of the particular NC surface chemistry. 

Polymers have been used widely in assembling colloidal networks.\cite{Zhao2012} For a surface-adsorbing polymer, this typically manifests in the bridging of particles to create a network because the same polymer chain can adsorb onto two different particle surfaces. While for a non-adsorbing polymer, an imbalance in osmotic pressure in the system leads to assembly via depletion attractions, covered in the following section. Thus, with increasing concentration of an adsorbing polymer to a particle dispersion, the following stages of phase behavior have been observed: dispersion $\xrightarrow{}$ bridging gel $\xrightarrow{}$ reentrant fluid $\xrightarrow{}$ depletion gel (Fig. 3a).\cite{Zhao2012} In other words, the same molecule can promote different phase behavior, acting as a bridging molecule or a depletant depending on its concentration in solution and the coverage or accessibility of particle surfaces. One example of this is the gelation of tin-doped indium oxide (ITO) NCs induced by low molecular weight polyethylene glycol (PEG) \cite{Cabezas2018}. A gel was observed at far lower concentration of PEG than predicted by Asakura Oosawa theory for depletion attractions. This gel was formed by adsorption of PEG on the acidic ITO surface via hydrogen bonding, leading to the bridging of NCs. More generally, when the number ratio of polymer chains (or other surface-adsorbing macromolecules) to NCs is less than the amount needed to saturate coverage of the NC surfaces, bridging may occur. The gelation regimes were accounted for by molecular thermodynamic theory to describe strong polymer-mediated associations between NCs, offering an avenue to predict which specific phase behaviors can be anticipated.

\begin{figure}
    \centering
    \includegraphics[width=.7\textwidth]{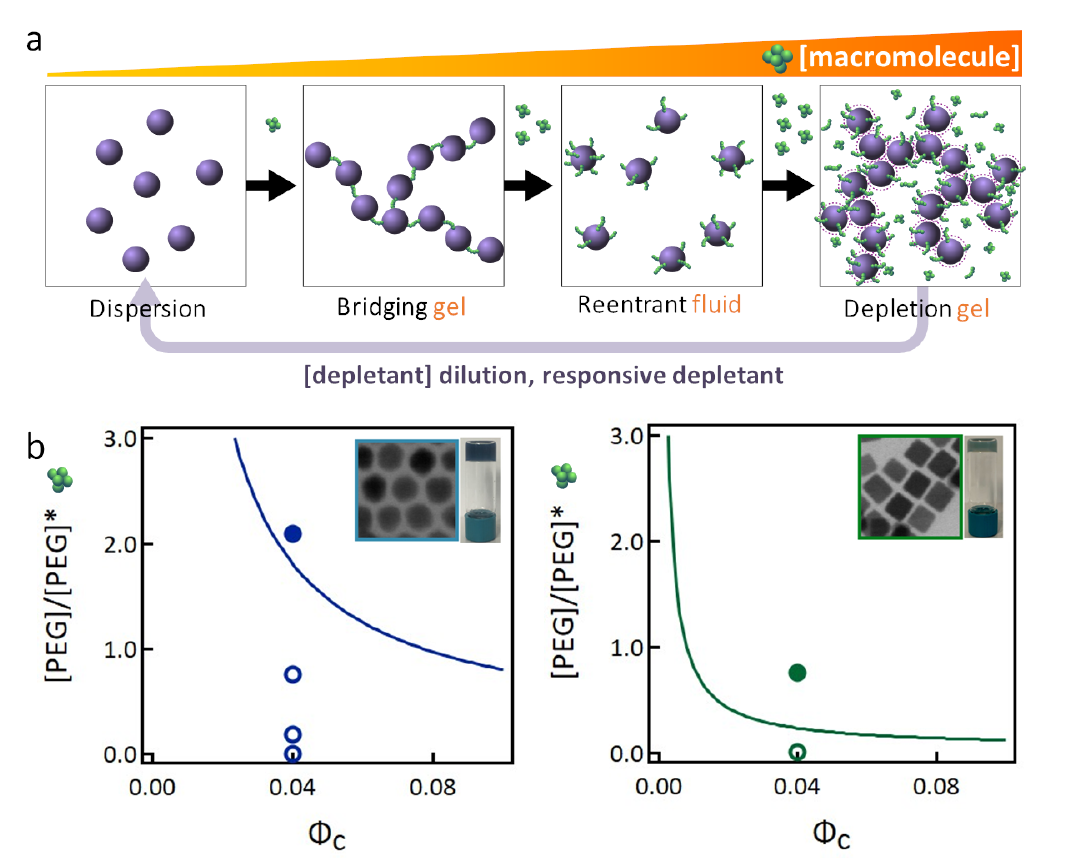}
    \caption{ (a) Schematic of NC phase behavior with increasing small molecule or polymer concentration: fluid $\xrightarrow{}$ bridging $\xrightarrow{}$ surface saturation $\xrightarrow{}$ depletion. (b) Phase diagrams of depletant volume fraction vs. NC volume fraction. Lines are calculated spinodal boundaries (effectively estimated gelation thresholds) and points are experimental data (open circle = fluid, closed circle = gel). Insets: digital photograph of NC gel in upside down vial and scanning transmission electron microscopy image of as-synthesized NCs. Adapted with permission from: (a) ref.~\citenum{Zhao2012}, Copyright 2012 Royal Society of Chemistry. (b) ref.~\citenum{Cabezas2020}, Copyright 2020 American Chemical Society. \label{fig:}}
\end{figure}

Similar bridging gelation and reentrant behavior was observed with the bridging of iron oxide (\ch{Fe_3 O_4}) NCs by four-arm nitrocatechol PEG ligands (4nPEG).\cite{Song2020} 4nPEG displaced a one-arm catechol PEG ligand originally on the NC surface; the 4nPEG has a greater binding affinity while still terminated with the same functional group to attach to the NC surface. At low 4nPEG concentrations, the system remained a dispersion, but as the amount of 4nPEG bridging molecule was increased, gelation occurred. Upon increasing the 4nPEG concentration further, stiffer and denser networks formed up until a threshold concentration. Beyond this threshold, reentrant softening occurred, followed by visible phase separation. Thus, by tuning the bridging molecule to NC ratio, one may control the phase behavior and the resulting mechanical properties.

The bridging gelation mechanism is straightforward to implement via selection of a bridging molecule that interacts with the NC surface. However, bridging induced aggregation may not generally be reversible, a consequence of the strong interaction between the bridging molecules and the NC surfaces. 
Nonetheless, this strategy offers a route to attain a gel with distinct properties from that of the dispersion. The strength of the gel and its corresponding properties may be tuned via the concentration of bridging molecule within a range before reentrant fluid behavior or precipitation occurs. Bridging gels formed with polymer bridges may offer additional tunability due to polymer conformational dependence on temperature (or other stimuli), which can modulate phase behavior. \cite{Haddadi2021} 

\section*{Depletion}

 While strongly adsorbing small molecules or polymers can act as bridges to aggregate NCs, even non-adsorbing molecules can have a similar effect due to depletion interactions.\cite{Bishop2009} Molecules that act as ``depletants'' in a NC dispersion cannot enter the overlapping exclusion region between closely spaced particles, creating an osmotic pressure imbalance that pushes neighboring NCs closer together (Fig. 3a).\cite{Asakura2014} The strength and range of this effective attraction can be controlled by adjusting the depletant concentration or its size relative to the NCs, so the depletion interaction depends solely on the physical properties of the depletants and the NCs.\cite{Lekkerkerker2011}. The ease of modulation via a secondary molecule and its universality across different composition NCs make depletion an attractive gelation mechanism.

Although depletion attractions have been used to assemble biomolecules or proteins\cite{Tanaka2002} and microparticles\cite{Lu2008}, only recently were such interactions leveraged for NC gelation. This was first demonstrated with the gelation of ITO NCs via polymer-induced depletion attractions, using low molecular weight PEG as the depletant \cite{Cabezas2018}. Shifting the balance from the long-range electrostatic repulsion of the charge-stabilized ITO particles to the short-range polymer-mediated attractions by adding PEG led to the formation of transparent, plasmonic metal oxide gels. When increasing the concentration of PEG at a fixed NC volume fraction, two gelation thresholds were observed, each preceded by a fluid regime. The first, discussed in the bridging gelation section above, was irreversible. However, because the strength of depletion attractions is proportional to the concentration of depletant, diluting the polymer readily reversed the second ``depletion gel'' back to a flowing dispersion. This reversibility, and the potential to reverse the corresponding structure-dependent properties, further highlights the tunability of gel assemblies formed via depletion interactions.

The strength of depletion attraction is proportional to the exclusion overlap volume between the larger particles that is inaccessible to depletant \cite{DelGado2016}.  Faceted particles experience stronger depletion attractions, when compared to spherical ones, due to their significantly larger overlap volumes. To demonstrate the universality of depletion attractions with respect to composition and shape, gelation studies were conducted with different compositions of NC spheres as well as cubes \cite{Cabezas2020}. Upon adding PEG to iron oxide (FeOx) NCs, ITO NCs and fluorine, tin co-doped indium oxide (FITO) NCs, the observed gelation threshold for spherical FeOx and ITO NCs agreed, while the cubic FITO NCs formed gels at a much lower [PEG], in agreement with theoretical phase diagrams (Fig. 3b). The same PEG concentration used to form the gel of cubic particles still resulted in flowing dispersions of spherical particles.

Depletion interactions as a route for NC gelation offer breadth in their applicability to different compositions, shapes, and sizes of NCs, as well as modularity in design by simply adjusting the size or concentration of the depletant to tune attractions. By avoiding dependence on the specific chemistries of NC surfaces or depletant molecules, this mechanism can be extended to a large variety of NC systems. With the aid of accurate theoretical predictions and phase diagrams, experimental discovery for future NC depletion gels may be accelerated as well.

\section*{Linking via coordination bonding}
In contrast to removing stabilizing ligands from NCs, gels can also be formed by adding linkers that form bridging bonds between ligands on neighboring NCs (Fig. 4a). 
Compared to the direct bridging of NCs, linking via coordination bonding offers more design opportunities to be generalized across different systems through the use of a linker that binds with functional ends of ligands, enabling reversibility and specificity. Linker-mediated NC gelation can be effectively controlled using covalent coordination bonding between bound ligands and metal ions (Fig. 4b). The optical, electronic, and magnetic properties of metal coordination complexes linking NCs can offer an additional means to introduce functionality in NC gels.\cite{Nag2012} Although various pairs of ligands and metal ions have been explored to form supramolecular assemblies and polymer gels, coordination bonding as a route to guide NCs gels has gained attention only recently.

Building blocks for coordination bonding-mediated gel networks can be prepared by functionalizing NCs with organic ligands terminated with complexing agents. For example, CdTe NCs functionalized with tetrazole-terminated ligands were synthesized by employing 5-mercaptomethyltetrazole as capping agent instead of conventional thioglycolic acid. \cite{Lesnyak2010} Upon addition of an aqueous solution of cadmium(II) acetate, the aqueous dispersion of CdTe NCs formed a transparent hydrogel within a short period of time (from a few seconds to 3 weeks, depending on the amount of cadmium(II) acetate added). Although the crystallinity of individual CdTe NC was not affected by gelation, the proximity of NCs within the gel led to the red-shift of photoluminescence (PL) maximum and significant reduction of emission lifetime. However, the NC dispersion was regenerated when a strong complexing agent, ethylenediaminetetraacetic acid (EDTA), was added, exhibiting partial recovery in optical properties (Fig. 4b). Similarly, \ch{Ni2P} nanoparticles capped with 11-mercaptoundecanoic acid formed a gel via coordination of ligand terminal groups (both carboxylate and thiolate) to nickel(II) ion. \cite{Hitihami2014} In this study, Ni$^{2+}$ was either externally added or dissolved out from NC surfaces that were rich in Ni$^{2+}$ after oxidation-induced thiolate decomplexation using tetranitromethane. The metal-assisted solvogels were then transformed into aerogels by supercritical drying, which had high surface area and thermal stability similar to aerogels prepared by chemical oxidation. 

Another novel gelation method was demonstrated using electrochemically generated metal ions. \cite{Hewa-Rahinduwage2021} CdS or CdSe quantum dots (QDs) functionalized with carboxylate-terminated ligands were crosslinked by metal ions (Ni$^{2+}$, Co$^{2+}$, Ag$^{+}$, or Zn$^{2+}$) generated by the anodic dissolution of a corresponding metal electrode. The QDs remained quantum-confined in gels, exhibiting similar structure, crystallite size, and optical properties to those of the QD building blocks. The gelation process was also chemically reversible using EDTA, confirming that the assembly of QDs was guided by metal ion-carboxylate coordination. The localized formation of gels using electrochemically generated metal ions exhibits potential to control the dimensions and shapes of gels more precisely compared to the conventional metal ion-mediated gelation methods. Recently, the competition between metal-mediated electrogelation and (the previously discussed) oxidative electrogelation was studied; it was found that the primary gelation mechanism was dependent on the oxidation potential, as well as the electrode material due to inhibition from electrogenerated by-products.\cite{Hewa-Rahinduwage2021B} These handles offer an avenue for tuning the dominant gelation mechanism and consequently the resulting gels.

More recently, coordination bonding was employed to induce thermoreversible gelation of ITO NCs, whose terpyridine-terminated ligands were interconnected via the addition of Co\textsuperscript{2+}.\cite{Kang2021} With increasing temperature in the presence of concentrated Cl\textsuperscript{-}, the linking bonds (cobalt(II)-(bis)terpyridine complexes) were gradually diminished in favor of the formation of cobalt(II) tetrachloride complexes, ultimately inducing phase transition of a gel into a free-flowing dispersion. The gel-to-sol transition at a threshold temperature was confirmed visually as well as via a decrease in scattering at low momentum vector ($q$) using \textit{in situ} temperature-dependent small-angle X-ray scattering (SAXS) (Fig. 4c). Because of the reversible nature of coordination bonding without irreversibly forming byproducts, the gelation process was highly cyclable and reproducible, which resulted in reversible shifting in the plasmonic peak positions of NCs during thermal cycling due to optical coupling in the gel state. The thermochromic properties of the metal coordination complexes enabled \textit{in situ} quantification of links during gelation, providing microscopic insight for the structural and thermodynamic factors controlling linker-mediated assembly. 

To avoid complications arising from insulating organic ligands and facilitate strong coupling between assembled NCs, all-inorganic NC gels were developed using coordination bonding between inorganic surface ligands and metal ions. For example, inspired by “chalcogels” made of chalcogenidometallate clusters (ChaMs) and Pt\textsuperscript{2+},\cite{Bag2007} NC gels were formed by adding Pt\textsuperscript{2+} to a dispersion of ChaM-capped CdSe NCs. Upon increasing concentration of Pt\textsuperscript{2+}, the NCs' phase transitioned from dispersion $\xrightarrow{}$ viscoelastic solvogels $\xrightarrow{}$ precipitates. \cite{Singh2015} The assembled microstructure was determined by SAXS, in which scattering intensity at low q increased to indicate degrees of clustering or aggregation. Notably, all-inorganic xerogels prepared by freeze-drying of these wet gels exhibited comparable surface area to previously reported aerogels produced by oxidation-induced gelation. 

Broadening the demonstration to all-inorganic NC gels of various compositions, the native organic ligands of a variety of NCs (CdSe, PbS, PbSe and ZnO nanospheres, CdSe nanoplatelets, and ZnO nanorods) were replaced by different inorganic ligands (S\textsuperscript{2-}, I\textsuperscript{-}, Cl\textsuperscript{-}, F\textsuperscript{-}, Ga/I and In/Cl complexes), and gelation was induced by adding optimal amounts of metal ions (Cd\textsuperscript{2+}, Pb\textsuperscript{2+}, Zn\textsuperscript{2+}) (Fig. 4d).\cite{Sayevich2016} However, due to the formation of strong covalent bonds directly bridging NCs, such as Se-Cd-Se, gelation was not reversible even when a strong complexation agent (EDTA) was introduced. Strong coupling between adjacent NCs in solvogels and aerogels was observed by broadening and peak shifting of absorption and PL spectra compared to that of the NC dispersion. Although not demonstrated yet, it is expected that all-inorganic NC gels can possess enhanced charge transport, magnetic and catalytic properties due to strong NC-NC coupling and distinctive properties of metal complexes bridging NCs. \cite{Nag2012}

\begin{figure}
    \centering
    \includegraphics[width=1\textwidth]{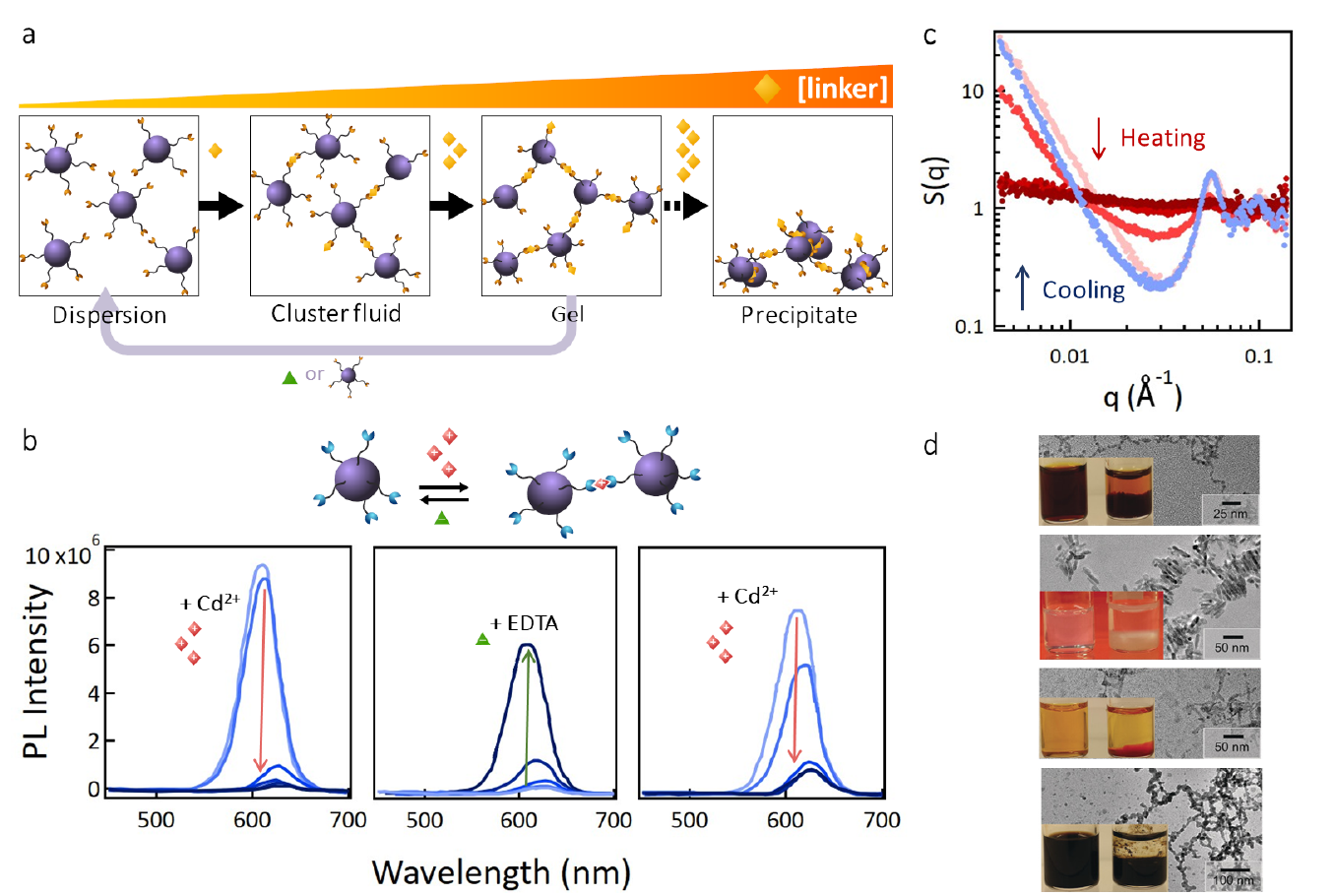}
    \caption{ (a) Schematic of NC assembly using bifunctional linkers (yellow) and subsequent disassembly either by using a capping agent (green) or by decreasing linker-to-NC ratio with additional NCs. (b) Schematic of linking NCs into a gel using coordination bonding and reversal with complexation agent (top) and PL response of CdTe NC dispersion upon adding Cd\textsuperscript{2+} and EDTA, sequentially (bottom). (c) SAXS structure factor of thermoreversible Co$^{2+}$ mediated NC gel, as a function of temperature. (d) TEM images of all-inorganic linker gels with insets of the corresponding NC dispersions (left) and gels (right). 
    Adapted with permission from: (a) ref.~\citenum{Dominguez2020}, Copyright 2020 American Chemical Society. (b) ref.~\citenum{Sayevich2016}, Copyright 2016 John Wiley and Sons. (c) ref.~\citenum{Kang2021}, Copyright 2021 The Authors. (d) ref.~\citenum{Lesnyak2010}, Copyright 2010 American Chemical Society.\label{fig:}}
\end{figure}

\section*{Linking via dynamic covalent bonding}
Similarly, bifunctional molecules that can form dynamic covalent bonds with their specific counterparts have emerged as promising linkers for programmable NC gelation. Especially, four sets of tunable, orthogonal, reversible, and covalent (TORC) pairs are of great interest because the orthogonality and reversibility of their reactions open the door to reconfigurable NC gels that are responsive to multiple chemical stimuli.\cite{Seifert2016, Reuther2018} The four TORC pairs include hydrazide/aldehyde, diol/boronic acid, thiol/conjugate acceptor, and metal/ligand. Ligands grafted on the surface of NCs are designed to display a functional terminal group that forms a dynamic covalent bond with a bifunctional linker. Although preparing ligands and linkers with specific functionality might be synthetically challenging, such modularity enables phase control of NCs regardless of their composition. In this strategy, the phase behavior of NCs can be controlled both microscopically and macroscopically, via rational design of ligands/linkers and by varying concentration of linkers, respectively (Fig. 4a).\cite{Sherman2021}

Dynamic covalent chemistry (DCC) allows the use of equilibrium principles to reversibly assemble NC gels, controlling the linker concentration to statistically impose valence restriction. The hydrazide/aldehyde pair was the first TORC pair leveraged for NC gelation.\cite{Dominguez2020} Specifically, ITO NCs were functionalized with aldehyde-terminated ligands, and oxalyldihydrazide was chosen as the linker. Via addition of the bifunctional linker above a threshold concentration, the NC dispersion transformed into a phase-separated gel in days to a few weeks, depending on linker concentration. The gelation was guided by the DCC between the aldehyde and hydrazide, leading to hydrazone formation. Interestingly, the amount of linker required to induce gelation was much greater than theoretically predicted due to the formation of self-loops, in which two ligands on a single NC are linked, not accounted for by Wertheim thermodynamic perturbation theory. \cite{Lindquist2016C} The discrepancy between theory and simulations, in which looping motifs are allowed, was resolved by relaxing this topological constraint in an updated and extended Wertheim theory. \cite{Howard2020A} Experimentally, network formation was monitored by SAXS, in which the diverging structure factor at low $q$ (< 0.03 \si{\angstrom}$^{-1}$) indicated the mesoscale organization of NCs due to significant attractions and consequent gelation. Due to coupling, gels showed significantly broadened plasmon absorption with reduced intensity compared to that of the NC dispersion. 
However, the infrared optical response was recovered when the gelation process was reversed either by adding a monofunctional capping agent or by reducing the linker-to-NC ratio with excess NCs. Notably, the same DCC pair was used to obtain binary nanoparticle aggregates, in which, without linkers, gold NPs decorated with complementary hydrazone groups were linked via a hydrazone exchange reaction. \cite{Marro2020} The aggregates were also reversed into colloidally stable clusters using a monofunctional aldehyde capping agent.

The reversibility and orthogonality of TORC chemistry provides an avenue to create programmable and reconfigurable NC ensembles, in which the phase behavior of different NC building blocks can, in principle, be controlled independently using distinct TORC pairs, or binary NC assembly with enforced local ordering (e.g. alternating compositions) can be obtained using asymmetric linkers. The modularity of the approach enables it to be readily adapted to new systems via substitution of NC building blocks while using the same molecular linkers to induce attractions.

\section*{Outlook}
To assemble NC gels, one must tune the inter-NC interactions either by the removal of existing repulsions or by the introduction of new attractions. With the former, investigators can use destabilizing molecules as a controlled method for removing NC surface ligands. The latter involves the addition of molecules to either bridge or link NCs, or the use of macromolecular depletants to induce entropic attractions. When the NCs are connected surface to surface, such as in the controlled destabilization or bridging mechanisms, their aggregation relies on their specific surface chemistry and is typically irreversible. However, through the use of secondary molecules indirectly interacting with the NCs (depletion, coordination bonding, DCC), one gains additional, reversible control over the strength of attractions leading to gelation. These key mechanistic principles can be used to aid the creation of new materials. With a target application or property in mind, the appropriate NC building blocks and gelation mechanism needs to be selected. This review laid out the main mechanisms used to assemble inorganic NC gels, each of which presents different opportunities for engineering functional materials.

With a better understanding of which intermolecular forces are at play and how NC aggregation proceeds towards gelation, one can more readily assemble networks of interest. Models have been developed to accurately describe kinetics of non-equilibrium, irreversible aggregation (progressive destabilization, bridging) as well as the thermodynamics of reversible aggregation based upon equilibrium principles.\cite{Jungblut2020} These models, when adopted in simulations or computed phase diagrams, help predict experimental conditions likely to result in phase behavior of interest. For reversible gelation methods (DCC, coordination bonding, depletion), strategies for network reconfiguration or disassembly can also be devised with the aid of computational tools.\cite{Sherman2021} Predicting and designing these assembled structures provides avenues for also designing the resulting, structure-dependent material properties. While arriving at the conditions for a specific property by experimental exploration alone would be time-consuming and resource-intensive, devising a feedback loop between experiments and theory instead may promote efficient design iterations toward targeted gel properties. 

Knowledge of the consequences or implications of each gelation strategy on the resulting gel characteristics can be leveraged to gain more precise control over structure-dependent properties. Understanding the underlying formation mechanism of each gel also presents the opportunity to employ stimuli-responsive components (linkers, depletants) to introduce even richer phase behavior and responsive properties. Although requiring the design of complex small-molecule mediators, the modularity of these approaches enables exchange of NC building blocks while leveraging the same gelation strategy alongside well-developed thermodynamic models. Dissipative assembly has also gathered recent interest due to the opportunity to access new, out-of-equilbrium structures with the addition of a stimulus or fuel before returning to the equilibrium state once the fuel has been consumed.\cite{Rao2021,Ravensteijn2020} This type of ``fueled'' assembly expands the possibilities for dyanamic NC gelation by using chemical and physical (e.g. light) additives to transiently activate or deactivate NCs, depletants, or linkers toward assembly and disassembly. Ultimately, the basis for engineering gels with desirable properties is understanding and controlling their formation mechanism. By thoughtfully assembling inorganic NC gels, scientists can leverage properties of the nanoscale world to create macroscopic materials that realize new functionality and applications.

\begin{acknowledgement}
This work was primarily supported by the National Science Foundation through the Center for Dynamics and Control of Materials: an NSF Materials Research Science and Engineering Center (NSF MRSEC) under Cooperative Agreement DMR-1720595. This work was also supported by the Welch Foundation (F-1848 and F-1696). E.V.A. acknowledges support from the Welch Regents Chair (F-0046).
\end{acknowledgement}

\bibliography{Overleaf_References}


\end{document}